\documentclass[doublecol]{epl2}

\usepackage{graphicx}
\usepackage{bm}
\usepackage{amsmath}
\usepackage{amssymb}
\usepackage{color}

\begin{document}
\newcommand{\erdosRenyi}{Erd\H{o}s-R\'{e}nyi }

\title{Synchronization in populations of sparsely connected pulse-coupled oscillators}

\author{A. Rothkegel}

\author{A. Rothkegel\inst{1,2,3} \and K. Lehnertz\inst{1,2,3}}
\shortauthor{A. Rothkegel \& K. Lehnertz}
\institute{
\inst{1}Department of Epileptology, University of Bonn, Germany\\
\inst{2}Helmholtz Institute for Radiation and Nuclear Physics, University of Bonn, Germany\\
\inst{3}Interdisciplinary Center for Complex Systems, University of Bonn, Germany
}


\abstract{We propose a population model for $\delta$-pulse-coupled oscillators with sparse connectivity. The model is given as an evolution equation for the phase density which take the form of a partial differential equation with a non-local term.  We discuss the existence and stability of stationary solutions and exemplify our approach for integrate-and-fire-like oscillators. While for strong couplings, the firing rate of stationary solutions diverges and solutions disappear, small couplings allow for partially synchronous states which emerge at a supercritical Andronov-Hopf bifurcation. }

\maketitle
The collective dynamics of interacting oscillatory systems has been studied in many different contexts in the natural and life sciences \cite{Winfree1967, Kuramoto1984, Pikovsky_Book2001, Arenas2008}. In the thermodynamic limit, evolution equations for the population density proved to be a useful description \cite{Desai1978, Omurtag2000, Acebron2005}, in particular to characterize the stability of synchronous and asynchronous states (see, e.g.,  \cite{Mirollo1990,Strogatz1991, Treves1993, Abbott1993, Strogatz2000,Vreeswijk2000, Gerstner2000,  Ly2010, Newhall2010, Louca2013}). Usually, dense or all-to-all-coupled networks are considered for these descriptions. Motivated by natural systems in which constituents interact with few others only, investigations of complex networks have revealed a large influence of the degree and sparseness of connectivity on network dynamics \cite{Hopfield1995, Golomb2000, Boergers2003,Zillmer2006, Zillmer2009, Rothkegel2011 , Luccioli2012,  Tessone2012}.  Especially 
when the knowledge about the connection structure is limited, it suggests itself to assume random connections (as in \erdosRenyi networks) or random interactions (where excitations are assigned randomly to target oscillators \cite{Omurtag2000, Sirovich2006, Dumont2013,Nicola2013}). Both approaches often yield comparable dynamics (e.g. \cite{Ferreira2012, Tattini2012}) whereas random interactions represents a substantial simplification from a mathematical point of view, allowing one to describe the networks in terms of evolution equations for the phase density. These equations are usually posed as starting point for the commonly applied mean- or the fluctuation-driven limits. However, rarely are they studied in full although it can be expected that sparseness largely influences the collective dynamics as has been discussed for excitable systems \cite{Sirovich2006}.

In this Letter, we propose a population model of $\delta$\nobreakdash-pulse coupled oscillators with sparse connectivity, derive the governing equations from a general definition of the density flux, and characterize existence and uniqueness of stationary solutions. For integrate-and-fire-like oscillators, the latter may either disappear with diverging firing rate or lose stability at a supercritical Andronov-Hopf bifurcation (AHB). This is in contrast to the global convergence to complete synchrony for all-to-all coupling that has been shown for finite~\cite{Mirollo1990} and for infinite~\cite{Mauroy2013} number of oscillators.

Consider a population of oscillators $n \in N$ with cyclic phases $\phi_n(t) \in [ 0,1 )$ and intrinsic dynamics $\dot{\phi}_n(t) = 1$. If for some $t_f$ and some oscillator $n$ the phase reaches 1, the oscillator fires and we introduce a phase jump in all oscillators $n'$ with probability $p= m/N$ \cite{DeVille2008, Olmi2010}. Here, $m$ is the number of recurrent connections per oscillator.  The height of the phase jump is defined by the phase response curve $\Delta(\phi)$ (PRC) (or equivalently by the phase transition curve $R(\phi)$):
 \begin{equation}\label{eq:interactionSingleOsci}
\phi_{n'}(t_f^+) = \phi_{n'}(t_f) +  \Delta\left(\phi_{n'}(t_f)\right) = R( \phi_{n'} ( t_f)).
\end{equation}
The model can be interpreted as an all-to-all coupled network in which connections are not 
reliable and mediate interactions between oscillators only with a small probability ($p$).  It can also be interpreted as an approximation to an \erdosRenyi network in which the quenched disorder, imposed by its construction, is replaced by a dynamic coupling structure which takes the form of an ongoing random influence.

For the limit of large sparse networks ($N \rightarrow \infty, m = \mbox{const.}$), we represent the network dynamics by a continuity equation for the phase density $\rho(\phi,t)$ 
\begin{equation}\label{eq:continuity}
\partial_t \rho(\phi,t) + \partial_\phi J(\phi,t) = 0
\end{equation}
with $\rho (\phi,t) \geq 0$ and $\int_0^1 \rho(\phi,t) d\phi = 1$. We assume the probability flux $J(\phi,t)$ to be continuous and define both $\rho$ and $J$ at phases $\phi \in [0,1)$. Evaluations at $\phi = 1$ are meant as left-sided limits towards $\phi = 1$. $J(1,t)$ is the firing rate. Every oscillator is subject to Poisson excitations $\eta_{\lambda(t)}$ with inhomogeneous rate $\lambda(t) = m J(1,t)$ and we can describe its phase variable by the stochastic differential equation $\partial_t \phi(t) = 1 + \eta_{\lambda(t)}$. To shorten our notation, we will omit in the following the time $t$ as argument of $\rho$, $\lambda$, and $J$. As we expect $R(\phi)$ to be non-invertible and to map intervals to a single phase, we have to take care in which way $\rho$ and $J$ are interpreted at these phases. 
Given some distribution of oscillators phases, we consider $\rho(\phi,t) d\phi$ as the fraction of oscillators which are contained in a small interval whose left boundary is fixed to $\phi$. With this definition, $\rho(\phi,t)$ is continuous for right-sided limits and the corresponding $J(\phi,t)$ is defined by the oscillators which pass an imaginary boundary which is infinitely close to $\phi$ and right to $\phi$. The flux can be formalized in the following way:
\begin {equation}\label{eq:fluxGeneralForm}
J(\phi) = \rho(\phi) + \lambda \left(\int \limits_{I_>(\phi)}  \rho(\tilde{\phi}) d\tilde{\phi}   - \int \limits_{I_\leq(\phi)}   \rho(\tilde{\phi}) d\tilde{\phi} \right),
\end{equation}
where $I_>(\phi)  := \{ \tilde{\phi} < \phi | R(\tilde{\phi}) > \phi\}$
is the set of phases smaller than $\phi$ which is mapped by $R(\phi)$ to a phase larger than~$\phi$, and $I_\leq(\phi) := \{ \tilde {\phi} > \phi | R(\tilde{\phi}) \leq\phi\} $ is defined analogously (the order relations in in these formulas are interpreted for unwrapped phases). The first term of the r.h.s. of \eqref{eq:fluxGeneralForm} represents convection due to the intrinsic dynamics of oscillators. The integrals represent the fractions of oscillators which are moved across phase $\phi$ by an excitation, either to smaller or larger values (cf. \eqref{eq:interactionSingleOsci}).  

PRCs which are derived from limit cycle oscillators by phase reduction usually have invertible phase transition curves \cite{Brown2004a}. However, \eqref{eq:fluxGeneralForm} even holds if $R(\phi)$ is not invertible and has no or uncountably many inverse images. For phases $\phi$ at which  $R(\phi)$ has at most countably many inverse images,  we can represent the sets $I_>(\phi)$ and $I_\leq(\phi)$ by a product of two Heaviside functions and derive, differentiating the latter to $\delta$-functions, the following expression:
\begin{equation}
	\partial_\phi J(\phi) = \partial_\phi \rho(\phi) + \lambda \int_0^1 \rho(\tilde{\phi})\left( \delta ( \phi - \tilde{\phi} )- \delta (\phi - R(\tilde{\phi}) \right) d\tilde{\phi}. 
\end{equation}
Denoting with $(R_i^{-1}(\phi) | i \in I)$ an enumeration of the inverse images of $R(\phi)$ at phase $\phi$ for an appropriate index set $I$, the continuity equation \eqref{eq:continuity} reads:
\begin{equation}\label{eq:sparseLimit}
	\partial_t \rho(\phi) = - \partial_\phi \rho(\phi) - \lambda  \rho(\phi)  + \lambda \sum_{i \in I} \frac{\rho(R_i^{-1}(\phi)) }{R'(R^{-1}(\phi))}.   
\end{equation}
For uncountably many inverse images of some phase $\varphi$, they will be contained in $I_>(\varphi)$ or $I_\leq (\varphi)$ but not in $I_>(\varphi^-)$ and $I_\leq(\varphi^-)$. In this case, we obtain a discontinuity between $\rho(\varphi^-)$ and $\rho(\varphi)$ which can be expressed by requiring continuity of the flux for left-sided limits at $\phi = \varphi$ ($J(\varphi^-) = J(\varphi)$). Note that the definition in \eqref{eq:fluxGeneralForm} automatically ensures continuity for right-sided limits. Setting $\phi = 1$ in \eqref{eq:fluxGeneralForm}, we obtain the following relationship for the excitation rate $\lambda= m J(1)$
\begin{equation}\label{eq:firingRate}
	\lambda = m \rho(1) /\left( 1 - m \int\limits_{I_>(1)} \rho(\tilde{\phi}) d\tilde{\phi}  + m \int\limits_{I_\leq(1)} \rho(\tilde{\phi})d \tilde{\phi}  \right).
\end{equation}
Given a PRC and the number of recurrent connections $m$, the population model for oscillators with sparse connectivity is given by \eqref{eq:sparseLimit}, \eqref{eq:firingRate}, and by the requirement that $\rho(\phi)$ is normalized to allow for an interpretation as probability density function. The integral over $I_\leq ( 1)$ in \eqref{eq:firingRate} corresponds to oscillators which pass the firing threshold in the \emph{wrong} direction. Usually, it is not desirable that such oscillators decrease the firing rate, which can be prevented by requiring the PRC to be bounded by $-\phi$ from below. Note that the excitation rate as defined in \eqref{eq:firingRate} may diverge or turn negative. In these cases every firing oscillator will make, on average, at least one other oscillator fire immediately and a macroscopic amount of oscillators fires in an instant. We will refer to this situation as an avalanche. Clearly, a numerical integration via some finite difference scheme will break down at this point \cite{
Kovacic2009, Dumont2013a}. Nevertheless, Monte-Carlo simulations may still be meaningful.

Let us briefly consider the mean-driven limit, i.e., a sequence of PRCs indexed by $i$ and parameters $m_i$ such that $\Delta_i(\phi)$ vanishes as $i \rightarrow \infty$ and the product  $\Delta_i(\phi) m_i$ converges point-wise to some function $Z(\phi)$.
The flux $J_i(\phi)$ is then straightforwardly approximated by
\begin{equation}\label{eq:denseLimitFlux}
	J_i(\phi) \rightarrow  \rho(\phi)  \left( 1 + J(1)  Z(\phi) \right)
 \end{equation}
as $i \rightarrow \infty$. Setting $\phi = 1$ in \eqref{eq:denseLimitFlux} gives the expression for the firing rate $\nu = \rho(1) / \left(1 - Z(1) \rho(1)\right)$ of the non-linear evolution equation 
\begin{equation}\label{eq:denseLimit}
	\partial_t \rho(\phi) = - \partial_\phi \left[ \left(1 + \frac {\rho(1)} {1 - Z(1)\rho(1)} Z(\phi)\right) \rho(\phi) \right ],
 \end{equation}
for which the continuity of the flux in \eqref{eq:denseLimitFlux} leads to the following non-linear boundary condition 
\begin{equation}\label{eq:denseLimitBoundary}
\rho(0) ( 1 + \lambda Z(0)) = \rho(1) (1 + \lambda Z(1)). 
\end{equation}
The dynamics of the system defined by \eqref{eq:denseLimit} and \eqref{eq:denseLimitBoundary} is easily describable for monotonous PRCs \cite{Mauroy2013}.  For increasing $Z(\phi)$ the probability density concentrates to a single phase in finite time for arbitrary initial distributions. For decreasing $Z(\phi)$ convergence to the stationary solution $\rho_0(\phi):= c/ ( 1 + c Z(\phi))$ can be observed. 

The question of synchronization in the population model with sparse connectivity can be addressed by investigating the existence and the stability of normalized stationary solutions of \eqref{eq:sparseLimit} and \eqref{eq:firingRate}. Stationary solutions of Eq.~\eqref{eq:sparseLimit} (with $\partial_t \rho(\phi,t) = 0$) can be obtained by segmenting $[0,1]$ into intervals in which oscillators receive either phase advances or retardations. For each of these intervals, a solution can then be obtained with some solver for delay differential equations with state dependent delays, stepping towards either larger or smaller phases. Phases $\varphi$ at which  $R(\phi)$ crosses the identity from below serve as suitable starting points for such a stepping approach because the sets $I_>(\varphi)$ and $I_\leq(\varphi)$ are empty at such points and we have $\rho(\varphi) = J(\varphi) = J(1) = \lambda /m$. Using this initial value, solutions fulfill \eqref{eq:firingRate}. In this way we can obtain stationary solutions $\rho(\phi;\lambda)$ of \eqref{eq:sparseLimit} and \eqref{eq:firingRate} in sole dependence on $\lambda$. Let us denote $I(\lambda) := \int_0^1 \rho(\phi;\lambda) d \phi$. In order to allow for a stochastic interpretation of $\rho(\phi;\lambda)$, $\lambda$ must then be chosen with shooting in such a way that $I(\lambda) = 1$. However, depending on the PRC such a choice may not be possible.  We can characterize the condition under which solutions exist by assuming that $R(\phi)$ is non-decreasing. Note that this assumption is valid for commonly considered PRCs including those of integrate-and-fire oscillators \cite{Brown2004a}. Under this assumption $I(\lambda)$ is strictly increasing in $\lambda$, which we will show at the end of this letter.  Stationary solutions are thus unique, and  to decide on their existence, it is thus sufficient to investigate the solutions $\rho(\phi;\lambda)$ for large $\lambda$. Near $\varphi$, the non-local term in \eqref{eq:sparseLimit} vanishes, and $\rho(\phi)$ decays as $e^{ - \lambda \phi} (\lambda / m)$ which has an integral independent on $\lambda$ and thus concentrates to $\delta(\phi - \varphi) / m$ for large $\lambda$. Analogously, $\delta$-peaks are generated at phases $R(\varphi), R^2(\varphi), ...$, at which oscillators arrive after having received a certain amount of excitations. Integrating over such a sequence of $\delta$-peaks, we can express $I(\infty)$ and thus characterize the existence of asynchronous solutions by the following inequality:
\begin{equation}\label{eq:existenceSolutions}
	I(\infty) = \max_i \{ i | R^{i} (\varphi) \leq 1 + \varphi \} / m > 1.
\end{equation}

\smallskip

We now report on findings of a dynamical analysis for the case of integrate-and-fire oscillators \cite{Peskin1975}. They are a popular model in many scientific fields ranging from physics and biology to the neurosciences \cite{Mirollo1990}. Our approach allows us to treat the non-invertible phase transition curves of excitatory and inhibitory oscillators and to study both dynamical regimes from a unified point of view. The PRC reads 
\begin{equation}\label{eq:PRCmirollo}
	\Delta(\phi) =  \max \{ \min \{ a\phi + b, 1 - \phi \}, -\phi \}. 
\end{equation}
The maximum and minimum bound $\Delta(\phi)$ by $-\phi$ from below and by $1 - \phi$ from above. The bound from above ensures that an excitation of oscillators cannot push them past the firing threshold  ($I_>(0) = \emptyset)$ and leads to uncountably many inverse images  $R(\phi)$ for $\phi = 1$. 
This assumption strongly favours synchronization; two oscillators adapt their phases completely after a suprathreshold excitation from one to the other.
The bound form below prevents oscillators with small phases which receive an inhibitory excitation to attain a negative phase or a phase just below the firing threshold ($I_\leq(1) = \emptyset$). We obtain the following boundary condition from \eqref{eq:fluxGeneralForm}:
\begin{equation}\label{eq:boundaryCondition}
	 \underbrace{\rho(0) - \lambda \int\limits_{I_\leq(0)} \rho(\tilde{\phi}) d\tilde{\phi}}_{J(0)}
	= \underbrace{\rho(1) + \lambda \int\limits_{I_>(1)} \rho(\tilde{\phi}) d\tilde{\phi}}_{J(1)}.
\end{equation}
The parameters $a$ and $b$ control both leakiness and coupling strength. For $a, b > 0$, $\Delta(\phi)$ represents excitatory integrate-and-fire oscillators  with concave-down charging function. 
For $a,b <0$ , $\Delta(\phi)$ represents inhibitory oscillators with concave-down charging function.
For other parameter combinations, $\Delta(\phi)$  represents excitatory and inhibitory oscillators with concave-up charging function and dynamical systems with both positive and negative phase responses.

\begin{figure}
\includegraphics[width = 0.95\columnwidth]{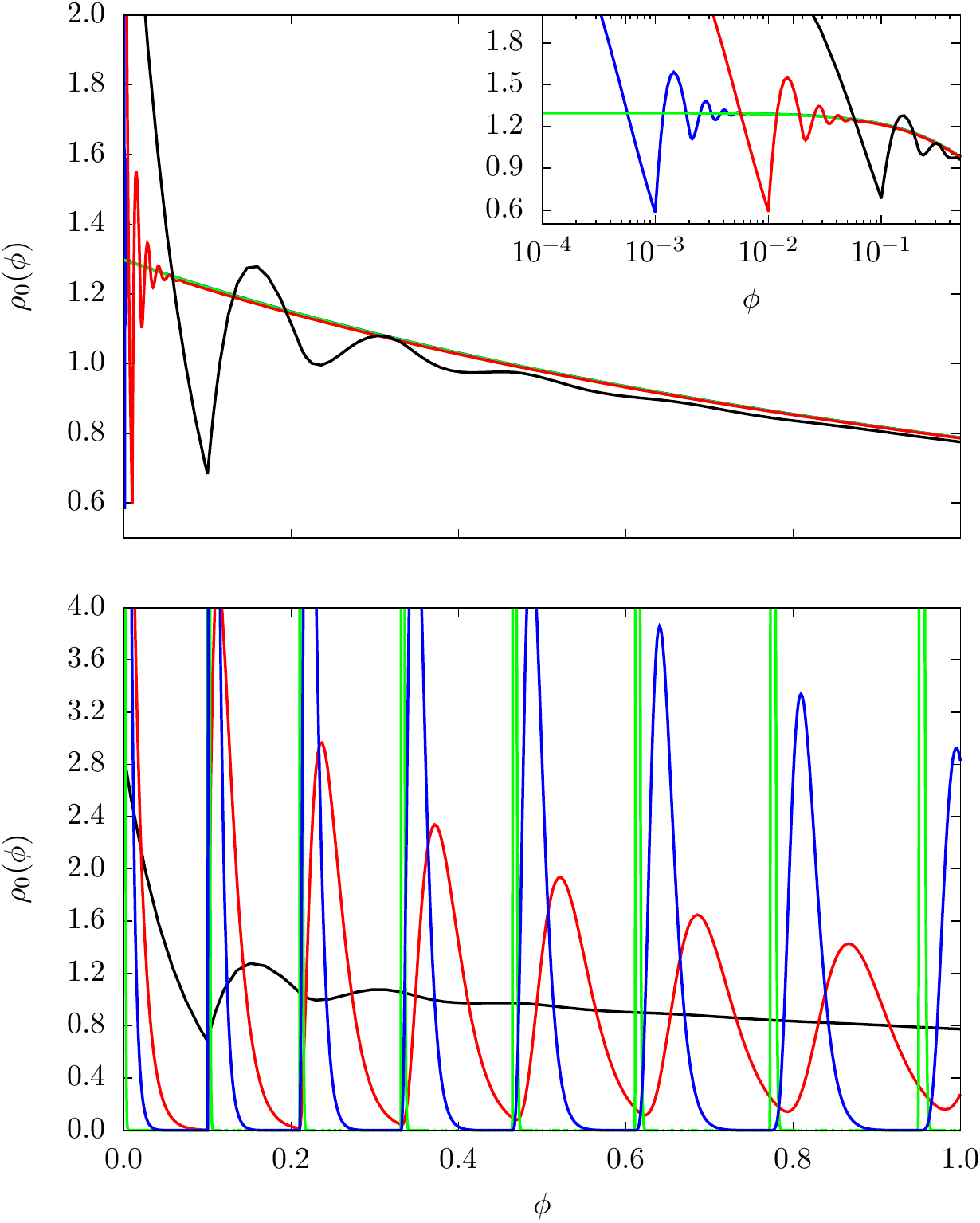}
\caption{(Colour on-line) Stationary solutions $\rho_0$  of \eqref{eq:sparseLimit} and \eqref{eq:firingRate} for the PRC given in \eqref{eq:PRCmirollo} with different parameter values $a, b$, and $m$ obtained by a shooting method with the constraint that $\rho_0$ is normalized.  Top: excitatory oscillators with $a = b = 0.5/m$, blue:  $m = 500$, red: $m=50$, black: $m = 5$. The green curve shows the stationary solution in the mean driven limit. Bottom: stationary solutions for $a = b = 0.1$ and different mean degrees $m$. Black: $m = 5.0$, red: $ m = 7.0$, blue: $m = 7.5$, green: $m = 8.0$. \label{fig:convergence}  }
\end{figure}

\begin{figure}
\includegraphics[width = 0.95\columnwidth]{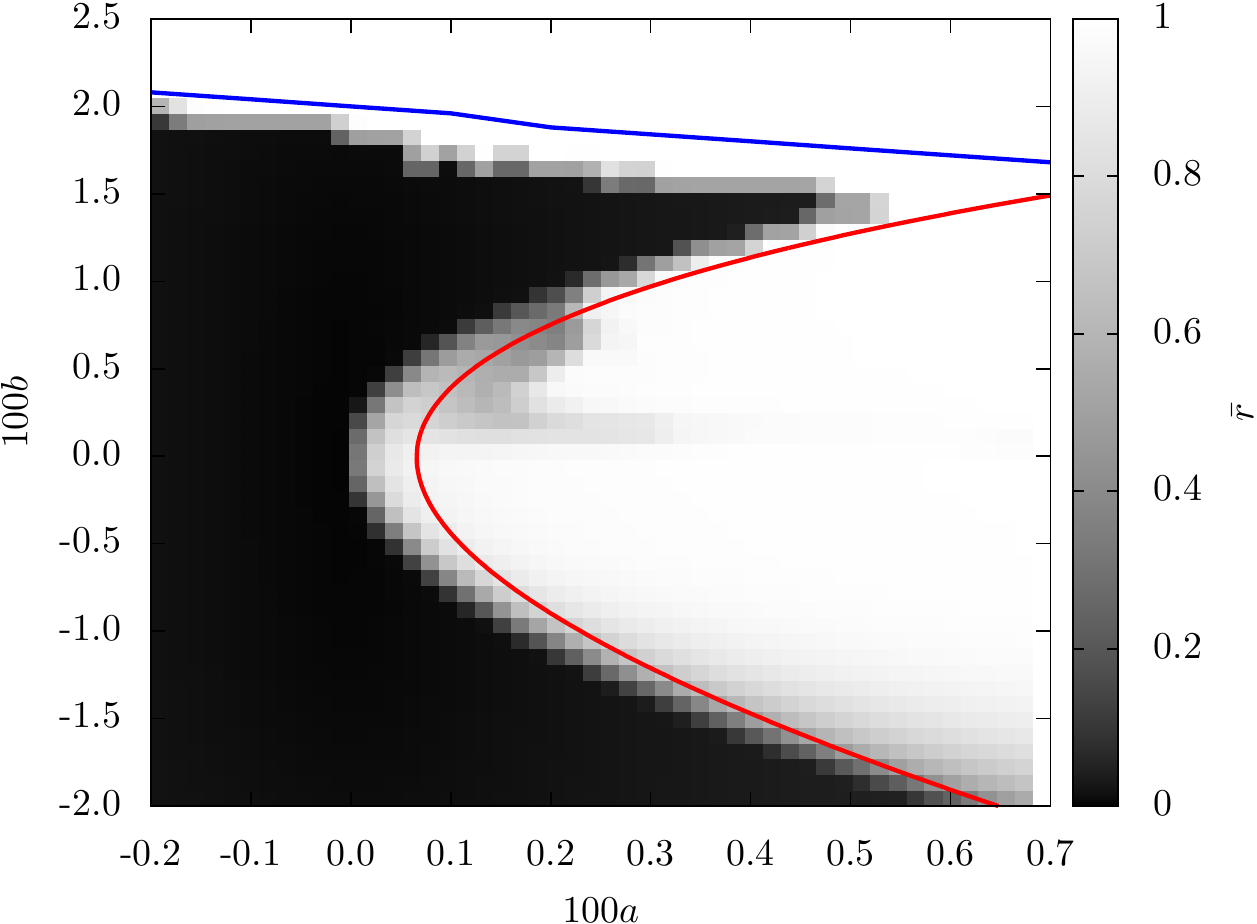}
\caption{(Colour on-line) Top: 
Mean order parameter $\bar{r}$ ($r(t):= 1 / \left|N\right| \sum_{n \in N} e^{2 \pi i \phi_n}$ averaged over $t\in [100,200]$) in dependence on $a$ and $b$ for a fixed mean degree $m = 50$.  Monte-Carlo simulations with $N = 10^6$ oscillators starting from uniformly distributed phases $\phi \in [0,1]$ \cite{Rothkegel2012}. The red line indicates the locus of supercritical AHB points obtained by AUTO \cite{Doedel97} after a finite difference discretization (2000 dimensions, first order upwind scheme) of \eqref{eq:sparseLimit} and \eqref{eq:firingRate}. The blue line indicates the upper boundary of the parameter regime in which stationary solutions can be obtained by shooting given by \eqref{eq:existenceSolutions}. \label{fig:parameterScan}}
\end{figure}

In Fig.~\ref{fig:convergence} we show stationary solutions for the PRC given in \eqref{eq:PRCmirollo}.

Given the PRC, phases in the interval $[0,b]$ are not reachable by excitations. The phase density $\rho(\phi)$ thus decays exponentially in this interval according to \eqref{eq:sparseLimit}. Phases $\phi > b$ are reachable, and $\rho(\phi)$ exhibits excursions of decreasing amplitudes, in which smoothed versions of the initial exponential segment in $[0,b]$ are repeated (cf.~\cite{Sirovich2006}).
Analogously, inhibitory oscillators with phases near $\phi=1$ do not receive excitations, which leads to a sharp decrease of $\rho(\phi)$ close to the firing threshold (not shown). For the mean-driven limit, stationary solutions show oscillations near $\phi = 0$ with frequencies that diverge with increasing $m$ (cf. Fig.~\ref{fig:convergence} top). For large coupling strengths or mean degrees, stationary solutions converge to a series of $\delta$-peaks (cf. Fig~\ref{fig:convergence} bottom) and eventually disappear.

For small $m$ and depending on oscillator parameters $a$ and $b$, 
we can distinguish different dynamics (cf. Fig.~\ref{fig:parameterScan}): asynchronous states with oscillator phases distributed according to the stationary solution, and (partially) synchronous states with oscillatory evolutions of the excitation rate $\lambda$. As for the mean-driven limit, large positive coupling strengths (above the black line in Fig.~\ref{fig:parameterScan}), do not allow for normalized stationary solutions with positive values which can be interpreted  as probability density. In this regime, oscillators synchronize completely within a few collective oscillations. Near this boundary the excitation rate diverges, and we observe no partially synchronous states.  For smaller coupling strengths, stationary solutions do exist and we now discuss their stability. 

We consider a small, localized perturbation of the stationary solution, which travels periodically around the phase circle.  An oscillator represented by this perturbation is shifted towards larger phase values due to its intrinsic dynamics and due to excitations. Both contributions are reflected by the corresponding terms in \eqref{eq:sparseLimit}.  The uncertainty of the oscillator's phase after some time leads to a broadening of the perturbation which increases with the strength of excitations and thus with $|b|$ (and to some degree with $a$).  When the perturbation crosses the firing threshold,  positive values of $a$ lead to a larger excitation of oscillators near the perturbation which leads to a sharpening.
Consequently, the perturbation vanishes for large $|b|$ and small $a$ and increases otherwise.
The boundary between both behaviors is characterized by a locus of Andronov-Hopf bifurcation (AHB) points. The AHB gives rise to oscillatory states with partial synchrony, in which a small perturbation of the stationary solution travels periodically around the phase circle. The amplitude of these oscillations increases with the distance to the AHB. For negative $b < a$, oscillators have negative phase responses near $\phi = 1$ and both integrals in the denominator in (\ref{eq:firingRate}) vanish. Consequently, we observe no avalanches and no phase concentrations in $\rho(\phi)$ up to some value of $b$ for which complete synchrony is reached.

For positive $a$ and $b$, the first integral in \eqref{eq:firingRate} does not vanish. When the amplitude of the oscillations grows so large that the denominator in \eqref{eq:firingRate} vanishes, an avalanche emerges. For larger values of $a$ 
, subsequent avalanches increase in size leading to complete synchrony after a few oscillations. For smaller values of $a$, these avalanches may, for finite networks, lead to complicated partially synchronous states with recurring avalanches, which, however, lie outside what can be described with the evolution equation. We will report on these states elsewhere (Rothkegel and Lehnertz, manuscript in preparation).

Note that the aforementioned broadening is not present in the mean-driven limit, in which both intrinsic dynamics and excitations are represented by a single convection term. If we consider the PRC in \eqref{eq:PRCmirollo} with parameters $a = \alpha / m$ and $b = \beta / m$, we obtain $Z(\phi) = \alpha \phi + \beta$ in the mean-driven limit ($m_i \rightarrow \infty$).  The phase density as determined by \eqref{eq:denseLimit} and \eqref{eq:denseLimitBoundary} thus 
converges to the stationary solution $\rho_0(\phi)$ for negative $a$ and concentrates for positive $a$ leading to complete synchrony of oscillators. In particular, the system does not allow for periodic solutions with partial synchrony of oscillators \cite{Mauroy2013}. In this case stable stationary solutions cannot be observed for $a > 0$.

We have presented a population model of $\delta$-pulse-coupled oscillators with sparse connectivity. Interactions between oscillators are defined by a phase response curve (PRC). We have defined the model in such a way that allowed us to treat non-invertible PRCs which lead to discontinuous distributions of oscillator phases. We have demonstrated the uniqueness of asynchronous solutions and characterized their existence. Finally, we have shown---using integrate-and-fire-like oscillators---two different mechanism which may lead to loss of asynchronous states. Stationary solutions may lose stability, giving rise to oscillations and partially synchronous states, or they may disappear completely, leading to avalanche-like synchronization and a fast convergence to synchrony.  We are confident that the model may further the understanding of the dynamics of sparsely coupled oscillatory networks. Systems that can be modelled as such appear ubiquitously in Nature. 

\smallskip

In this last part of the letter, we will show that $I(\lambda)$, the norm of stationary solutions of (5) and (6), is strictly increasing in $\lambda$, provided that the phase transition curve $R(\phi)$ is increasing in $\phi$ and crosses the identity at one or more points from below. For the sake of simplicity, we will shift the phases in such a way, that the crossing occurs at $\phi = 0$ such that we have $\rho(0) = J(0) = \lambda / m$. As first step, we relate $I(\lambda)$ defined for some PRC to an exit-time problem for the stochastic dynamics of oscillators  $\partial_t \phi(t) = 1 + \eta_{\lambda(t)}$ which is determined by convection with velocity 1 and by Poissonian excitations $\eta_{\lambda(t)}$ with inhomogeneous rate $\lambda(t)$.
To this end, we consider the interval $(0,1)$ to be empty at $t = 0$. If we now inject a constant flux $J(0)$, oscillators will pass the interval and exit at $\phi = 1$ after some variable time $t_\mathrm {E}$. We consider the distribution $P(t_\mathrm{E})$ of these exit times. The flux $J(1,t)$ will increase from $J(1,0) = 0$ and will eventually approach the injected amount $J(0)$, at which time the same amount of oscillators enter and exit the interval. If we inject $J(0) = \lambda /m$, then the number of oscillators which are in the interval at a large time $t$, is given by $I(\lambda)$ and can be expressed by integrating over the difference of incoming and outgoing fluxes:
\begin{equation}
	I(\lambda) = \lim_{\kappa \rightarrow \infty} \int_0^\kappa \frac {\lambda}{m} -    J(1,t) dt.
\end{equation}
Given the distribution of exit times $P(t_\mathrm{E})$, we can express the outgoing flux by integrating over the time $t_0$ at which oscillators are injected into the interval:
\begin{equation}
	I(\lambda) =  \lim_{\kappa \rightarrow \infty} \left(\frac{\lambda}{m} \kappa -  \int_0^\kappa dt  \int_0^t dt_0   P(t - t_0) \frac{\lambda}{m} \right). 
\end{equation}
The domain of the integral is the area of the quadrant $(t_0, t) \in [0,\kappa] \times [0,\kappa]$  which lies above the diagonal. If we parametrise this domain by $t_\mathrm{E}:= t - t_0$ and $t_0$, we obtain, using substitution for multiple variables,
\begin{equation}
	I(\lambda) =  \frac{\lambda}{m} \lim_{\kappa \rightarrow \infty} \left( \kappa  - \int_0^\kappa dt_\mathrm{E} \int_0^{\kappa - t_\mathrm{E}} dt_0  P(t_\mathrm{E})  \right ).
\end{equation}
As every oscillator eventually reaches $\phi = 1$, we have $\int_0^\infty P(t_\mathrm{E}) dt_\mathrm{E} = 1$, and we obtain a surprisingly simple relationship, which says that the norm of $I(\lambda)$ is given by the product of the injected flux and the mean exit time: 
\begin{equation}\label{eq:increasingI}
	I(\lambda)	= \frac{\lambda}{m} \int_0^\infty { t_\mathrm{E} P(t_\mathrm{E}) d t_\mathrm{E} }.
\end{equation}
Let us represent Eq.~\eqref{eq:increasingI} by an integral equation. We define $M(\varphi)$ as the mean time an oscillator with phase $1 - \varphi$ remains in the unit interval before it reaches $\phi = 1$. With this definition, the mean exit time for the entire interval is $M(1)$. Oscillators outside of the interval have a vanishing exit time: $M(\varphi) = 0$ for $\varphi < 0$. $M(\varphi)$ can now be expressed by an average over the time of the next excitation. Assuming an exponential distribution $\iota(t) := \lambda e^{-\lambda t}$ for the times between excitations, we can relate these times to probabilities. With probability $\bar{\iota}(\varphi) = 1 - \int_0^\varphi \iota(t) dt$, the oscillator will leave the interval without receiving another excitation. For the case that the oscillator receives an excitation at time $t$ after injection, it has a phase of $1-\varphi + t + \Delta(1 - \varphi + t)$ afterwards. The mean time the oscillator needs to pass the remaining phase distance can again be expressed by $M(\varphi)$ 
which results in the following integral equation for $M(\varphi)$:
\begin{equation}\label {eq:volterraM}
	M(\varphi) =  \bar{\iota}(\varphi) \varphi  + \int_0^\varphi \iota(t) \left[ t  + M(\varphi - t - \Delta(1 - \varphi + t)) \right ] dt .
\end{equation}
Inserting $\iota(t)$  and multiplying Eq.~\eqref{eq:volterraM} by $\lambda / m$, we obtain a similar equation for $I(\lambda,\varphi) : = \int_0^\varphi \rho(\phi;\lambda/m,\lambda)$ which we define as generalization of $I(\lambda)$ with $I(\lambda,1) = I(\lambda)$:
\begin{equation}  \label{eq:volterraMeanExitTimes2}
	I (\lambda, \varphi)	 =  \frac{1 - e^{-\lambda \varphi}}{m} + \int_0^\varphi   \lambda e^{-\lambda(\varphi -t)}  I (\lambda, t- \Delta(1-t)) dt.
\end{equation}
For convenience, we will use the abbreviations $G(\lambda,\varphi):= (1- e^{-\lambda \varphi}) / m$ and $z(\varphi, u) := \varphi  + u - \Delta(1 - \varphi - u)$. $G(\lambda,\varphi)$ is strictly increasing in both arguments. Using our assumption about the PRC, we see that $z(\varphi,u)$ is increasing in both arguments but not necessarily strictly increasing. Eq.~\eqref{eq:volterraMeanExitTimes2} takes the following form:
\begin{equation}  \label{eq:volterraMeanExitTime3}
	I (\lambda, \varphi)	 =  G(\lambda, \varphi) + \int_0^\infty \lambda e^{-\lambda u} I (\lambda, z ( \varphi, -u)) du.
\end{equation}
We have extended the integral from $\varphi$ to $\infty$  using  $I(\lambda,\varphi) = 0$ for $\varphi  < 0$. The equation is a Volterra integral equation of the second kind. 
Note that $z(\varphi, -u) \leq \varphi$ such that the equation defines $I(\lambda, \varphi)$ in a hierarchical way. $I(\lambda, \varphi_0)$ is obtained by taking some value $G(\lambda, \varphi_0)$ and by adding an weighted average over previous values $I(\lambda, \varphi), \varphi < \varphi_0$. We can thus conclude  that $I(\lambda,\varphi)$ is positive, if $G(\lambda,\varphi)$ is positive for all $\varphi$.

We demonstrate that $I(\lambda)$ is strictly increasing in $\lambda$ for PRCs with $\partial_\phi R(\phi) \geq 0$.  We first argue that $I(\lambda,\varphi)$ is strictly increasing in its second argument for every $\lambda > 0$. Differentiating \eqref{eq:volterraMeanExitTime3} by $\varphi$, we obtain an integral equation for $\partial_\varphi I(\lambda, \varphi)$ which is of the same kind as \eqref{eq:volterraMeanExitTime3} and has a non-negative kernel $\lambda e^{-\lambda u} \partial_\varphi z(\varphi, -u)$ and a positive function $\partial_\varphi G(\lambda, \varphi)$ outside of the integral. Analogously, we can thus conclude that $\partial_\varphi I(\lambda, \varphi)$ is positive. Finally, we argue that $I(\lambda, \varphi) $ is increasing in $\lambda$ for every $\varphi \leq 1$. Taking the derivative of \eqref{eq:volterraMeanExitTime3} with respect to  $\lambda$, we obtain three terms according to the dependences on $\lambda$ of $G$, of the kernel, and of $I$. The first term $\partial_\lambda G(\lambda, \varphi)$ is positive as G is strictly increasing in its arguments. As second term, we obtain 
\begin{equation}
\int_0^\infty I(\lambda, z(\varphi, -u)) \partial_\lambda \lambda e^{-\lambda u} du.
\end{equation}
Using $\partial_\lambda  (\lambda e^{-\lambda u}) = \partial_u ( u e^{-\lambda u})$ and integrating by parts, we obtain also a positive contribution as $I$ and $z$ are increasing in their second arguments. The third term contains a weighted average via a positive kernel of $\partial_\lambda I(\lambda, \varphi)$. As before, we infer that $\partial_\lambda I(\lambda,\varphi)$ is strictly increasing which gives for $\varphi = 1$ the desired proposition.

We are grateful to Stefano Cardanobile for fruitful discussions and Gerrit Ansmann for careful revision of an earlier version of the manuscript. This work was supported by the Deutsche Forschungsgemeinschaft (LE 660/4-2).


\end{document}